# String Phase in an Artificial Spin Ice

Xiaoyu Zhang[1,2,3], Ayhan Duzgun[4], Yuyang Lao[2,3], Shayaan Subzwari[1], Nicholas S. Bingham[1], Joseph Sklenar[2,3,5], Hilal Saglam[1], Justin Ramberger[6], Joseph T. Batley[6], Justin D. Watts[6,7] Daniel Bromley[8], Rajesh V. Chopdekar[9], Liam O'Brien[8], Chris Leighton[6], Cristiano Nisoli[4], & Peter Schiffer[1,2,3,10 *]

[1]*Department of Applied Physics, Yale University, New Haven, CT 06511, USA*

[2]*Department of Physics, University of Illinois at Urbana-Champaign, Urbana, Illinois 61801, USA*

[3]*Frederick Seitz Materials Research Laboratory, University of Illinois at Urbana-Champaign, Urbana, Illinois 61801, USA*

[4]*Theoretical Division and Center for Nonlinear Studies, MS B258, Los Alamos National Laboratory, Los Alamos, New Mexico 87545, USA*

[5]*Department of Physics and Astronomy, Wayne State University, Detroit, Michigan 48201, USA*

[6]*Department of Chemical Engineering and Materials Science, University of Minnesota, Minneapolis, Minnesota 55455, USA*

[7]*School of Physics and Astronomy, University of Minnesota, Minneapolis, Minnesota 55455, USA*

[8]*Department of Physics, University of Liverpool, Liverpool L69 3BX, United Kingdom*

[9]*Advanced Light Source, Lawrence Berkeley National Laboratory, Berkeley, CA 94720, USA*

[10]*Department of Physics, Yale University, New Haven, CT 06511, USA*

* Corresponding author email: peter.schiffer@yale.edu

**Abstract**

One-dimensional strings of local excitations are a fascinating feature of the physical behavior of strongly correlated topological quantum matter. Here we study strings of local excitations in a *classical* system of interacting nanomagnets, the Santa Fe Ice geometry of artificial spin ice. We measured the moment configuration of the nanomagnets, both after annealing near the ferromagnetic Curie point and in a thermally dynamic state. While the Santa Fe Ice lattice structure is complex, we demonstrate that its disordered magnetic state is naturally described within a framework of emergent strings. We show experimentally that the string length follows a simple Boltzmann distribution with an energy scale that is associated with the system's magnetic interactions and is consistent with theoretical predictions. The results demonstrate that string descriptions and associated topological characteristics are not unique to quantum models but can also provide a simplifying description of complex classical systems with non-trivial frustration.

## Introduction

Numerous exotic phenomena arise in strongly correlated many-body systems, even when the underlying interactions are simple, and artificial spin ice arrays composed of coupled single-domain nanomagnets are an important class of such systems[1,2,3]. While artificial spin ice studies originally focused on connections to the frustration-induced phenomena seen in pyrochlore spin ice materials[4], such as monopole-like excitations[5], the field has now expanded to include a wide range of exotic behavior in carefully designed geometries[6,7,8,9,10,11,12,13,14,15,16]. We have experimentally studied the Santa Fe Ice geometry of artificial spin ice,[17,18], demonstrating that the local excitations among the nanomagnet moments are correlated in Boltzmann-distributed one-dimensional strings. One-dimensional strings of local excitations are an important characteristic in strongly correlated topological quantum matter[19,20,21,22,23,24], and our data demonstrate that such strings can also be observed in a classical thermal system.

## Results

The structure of Santa Fe Ice (SFI)[17,18] is shown in Figure 1a, where each island is a single-domain nanoscale ferromagnet that behaves like a binary Ising-like moment. Note that the structure of SFI, while somewhat complex at first sight, is obtained in a straightforward manner by removal of a subset of moments from the simple square lattice[17]. Figure 1b shows a scanning electron microscope image of one of our experimental samples, and Figures 1c and 1d show experimental measurements of the individual moments in SFI, as described in detail below. The unit cell of SFI, indicated in Figure 1a, is composed of two composite large squares, each composed of eight elementary rectangular plaquettes consisting of six moments – two interior plaquettes (shaded in yellow) and six peripheral plaquettes (unshaded).

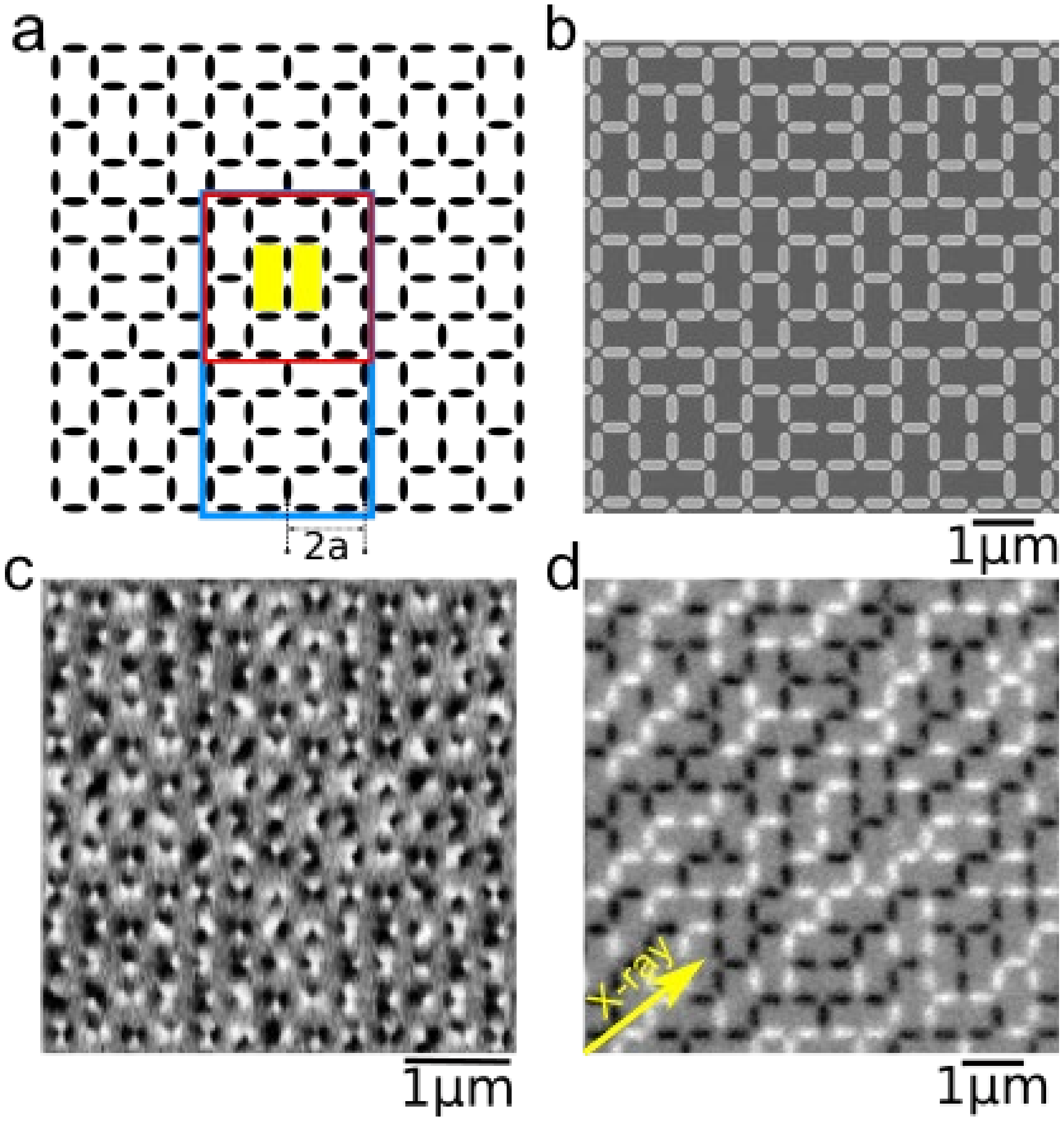

**Figure 1: Santa Fe ice. (a)** Schematic of the Santa Fe Ice (SFI) geometry, where each element represents a single-domain nanomagnet, and the lattice constant is *a*. The unit cell (indicated in blue) is made of two composite squares (one of which is indicated in red). Each composite square has eight rectangular plaquettes, which can be categorized as pairs of “interior” plaquettes (indicated by yellow shading) that are separated by a pair of islands, and “peripheral” plaquettes that surround them. **(b)** Scanning Electron Microscopy (SEM) image of SFI. (**c**). Magnetic force microscopy (MFM) image of SFI, in which each ferromagnetic island has black and white contrasting ends, indicating the moment poles. **(d)** X-ray magnetic circular dichroism photoemission electron microscopy (XMCD-PEEM) image of SFI, in which the entire islands are either black or white,

indicating the magnetic moment direction through its component projected onto the incident X-ray beam (yellow arrow).

We first analyze the structure of SFI to establish a framework for the analysis of our experimental data. Following previous analyses of artificial spin ice systems,[1,2,3] we describe SFI in a near-neighbor approximation via a vertex model, i.e., we consider the system through the states of the lattice vertices, as defined by the possible configurations of moments at each vertex of the lattice (Figure 2a). SFI belongs to the class of artificial spin ices that are "vertex frustrated" [17,18]. Rather than the individual magnetic moments being frustrated in their interactions, as is typical for geometrically frustrated magnets, frustration arises in the spatial allocation of low-energy vertex configurations. In other words, the moments cannot be arranged such that each vertex is in its local ground state (the ground state for the vertex moments when considering just the interactions within the vertex), because the lattice structure of SFI forces some fraction of the vertices to be in a local excited state. These excited vertices have been dubbed "unhappy vertices" in previous works[17,25], contrasting with "happy vertices" that are in their local ground states.

Within this vertex framework, the proper description of the disorder in SFI is not in terms of magnetic moments but rather in terms of the degenerate allocation of unhappy vertices and of the rules controlling their constrained disorder. In Figure 2a, we show all possible moment configurations on each vertex type, noting that vertices in this lattice can include either two, three, or four moments, i.e., they can have a coordination number of $z$ = 2, 3, or 4. We describe two plaquettes bordered by an unhappy vertex as being "connected" with each other by the unhappy vertex, as illustrated in Figure 2b where the connected plaquettes are each joined by a blue line drawn through the pair of moments whose relative configuration differs from that in a vertex ground state.[8,26]

Since the two moments in the center of each composite square form a $z = 2$ vertex, within the particular structure of SFI, each of the interior plaquettes must be connected by an *odd* number of unhappy vertices around its edges. Those plaquettes are therefore intrinsically frustrated in that they cannot have all vertices in a ground state. Similarly, each of the peripheral plaquettes must be connected by an *even* number of unhappy vertices (including possibly zero). These two conditions, illustrated and explained in detail in the Supplementary Note 2, can be considered as topological constraints: there must be at least one unhappy vertex connected with each interior plaquette, and there could be zero or any even number connected with each peripheral plaquette.

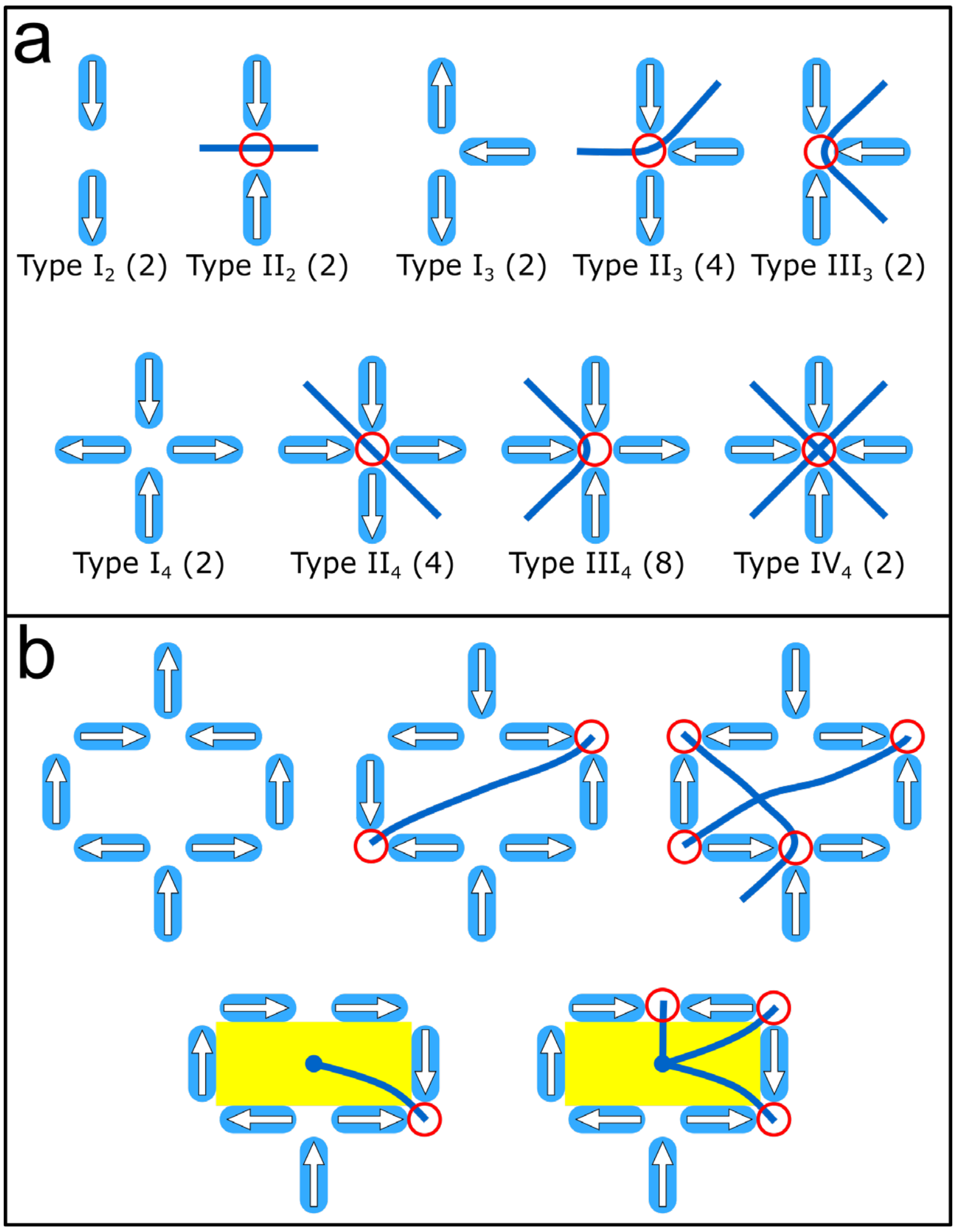

**Figure 2: Vertex types and string representation. (a)** Vertex moment configurations for different vertex coordination numbers, $z$ = 2, 3, and 4, arranged in order of increasing energy, where the arrows indicate moment direction and the red circles denote an excited state of the vertex (i.e., an unhappy vertex). The numbers in parentheses show the degeneracy for each given vertex type. The blue lines represent string segments “connecting” the plaquettes through the unhappy vertices. **(b)** Examples of moment and string configurations for peripheral (unshaded) and interior (yellow-shaded) plaquettes. Because of the SFI lattice geometry, a peripheral plaquette is connected by an even number of unhappy vertices, and an interior plaquette is connected by an odd number.

The consequences of viewing SFI through the connected plaquettes are striking. Consider an unhappy vertex that connects an interior plaquette to a peripheral plaquette. In order to yield an even number of unhappy vertices for that peripheral plaquette, a second unhappy vertex must also be connected to it. That vertex in turn connects to another peripheral plaquette, and it is thus natural to attach the unhappy vertices together and visualize their connectivity as one-dimensional strings within the lattice[18]. If we put line segments through each unhappy vertex connecting those plaquettes, as in Figure 2, and then attach the segments, we find that all the unhappy vertices can be represented as belonging to such strings, with the string length measured as the number of connected unhappy vertices. These strings typically only start and end in the interior plaquettes, because of the odd number of connecting vertices in those plaquettes, although loops that only include peripheral plaquettes are possible at higher temperatures (see Supplementary Fig. 3).

Since the strings represent topological constraints of the SFI moment ensemble, this string representation is valid at any temperature, but it is particularly useful in considering low-energy states of the system. In fact, Monte Carlo simulations demonstrate that the two possible ground states of the moment configuration are characterized by their string configurations (see Supplementary Note 1). For larger next-nearest-neighbor interactions relative to nearest-neighbor interactions, the ground state is long-range-ordered, with strings of length one connecting pairs of contiguous interior plaquettes. In the limit of larger nearest-neighbor interactions, the ground state is characterized by a disordered and highly degenerate state of strings connecting separated interior plaquettes. Our experiments probe this latter regime, as demonstrated in the data below.

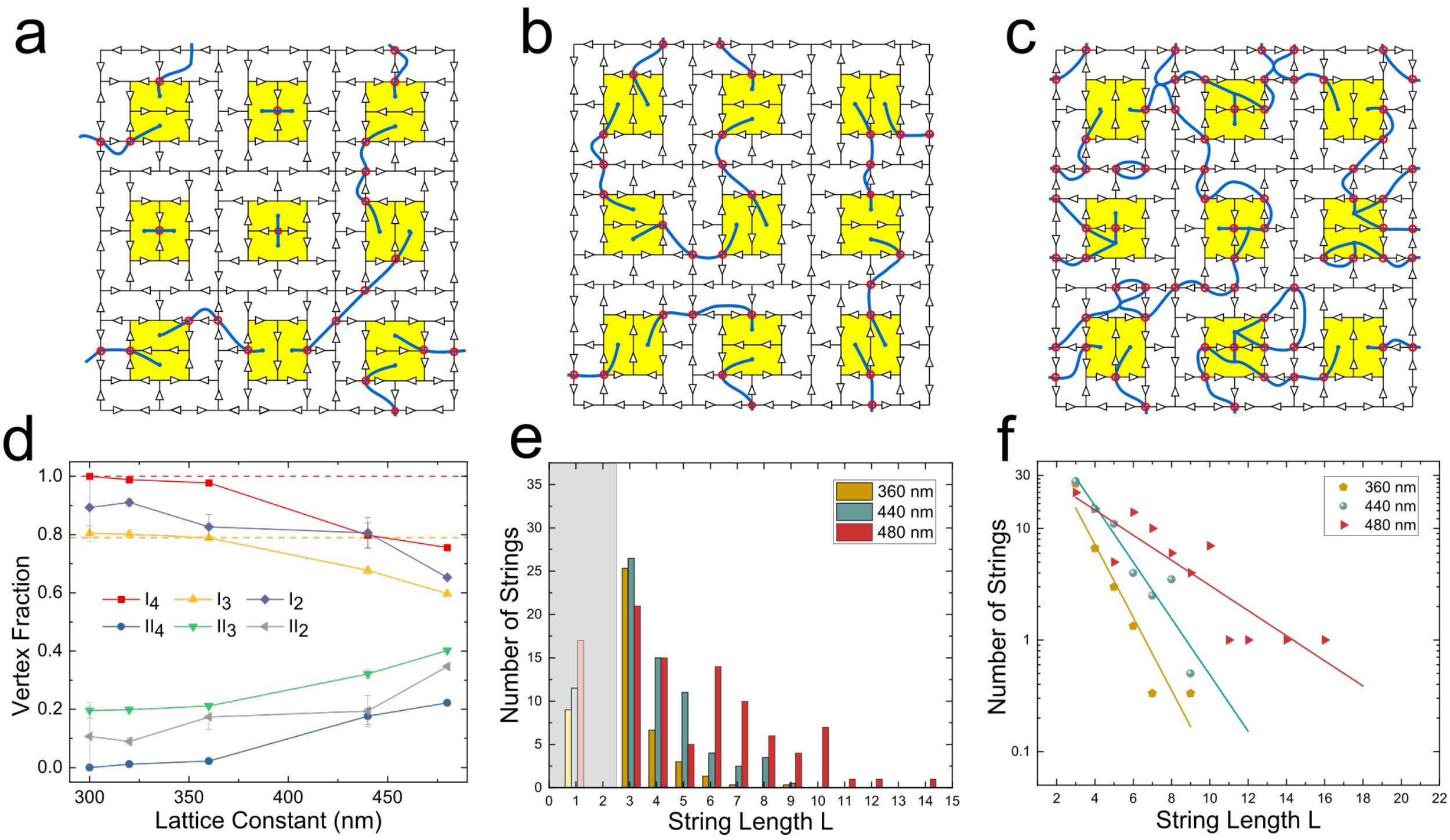

**Figure 3: Thermal annealing experimental results. (a-c)** Examples of moment configurations for small portions of the arrays measured after annealing, with both the moments and strings represented for samples with lattice constants of 320 nm, 360 nm, and 480 nm, respectively. Additional raw MFM data are given in Supplementary Note 3. **(d)** The vertex fractions observed after annealing as a function of lattice spacing (error bars represent standard deviations of the data). With decreasing lattice spacing, i.e., increasing interactions, the fractions converge to those expected for the disordered string ground state. Dashed lines are the expected vertex fraction of $I_4$ and $I_3$ for disordered string ground state. **(e)** The distribution of string lengths for different lattice spacings where the grey section covers those strings that are connecting only contiguous interior plaquettes and therefore are expected to have a different energy scale. We estimate the uncertainty in these values to be approximately 10-15%, based on the number of strings in the images. **(f)** The data from (e) plotted on a semi-logarithmic scale with exponential fits as described in the text.

We now turn to the experimental behavior of SFI, as viewed within the framework of unhappy vertices being considered collectively as correlated strings. We studied permalloy ($Ni_{80}Fe_{20}$) SFI both in a *static* configuration after high temperature annealing and in a *dynamic* state undergoing thermally induced moment reversals. In these two distinct cases, which correspond to thicker and thinner permalloy islands, we image the moment configurations through magnetic force microscopy (MFM) and x-ray magnetic circular dichroism photoemission electron microscopy (XMCD-PEEM), respectively. Details of island size, lattice constant, sample fabrication and measurement protocols are given in the Methods section, and representative images of the moment configurations obtained with each technique are shown in Figures 1c and 1d. In both cases we characterize the fractions of the different vertex states and we extract the distribution of string lengths, *L*, using graph-theoretical techniques described in the Supplementary Note 5.

Thermal annealing of the thicker islands is very efficient in reaching low energy states[6,27,28,29] because it starts near the Curie point, at which the permalloy has reduced magnetization. Thus, the constraints predicated upon the binary nature of the island magnetization necessarily break down during this process. Figures 3a – 3c show typical real space snapshots obtained via MFM of annealed samples for different lattice spacings, demonstrating the different sorts of string configurations observed. Figure 3d then plots the vertex statistics of annealed samples as a function of their lattice constants. For strongly coupled samples, i.e., small lattice spacings, the data converge to the fractions expected for the disordered string ground state, where each string consists of three excited three-island vertices out of fourteen. The dashed lines represent the expected vertex fraction. Figure 3e and 3f show the distribution of string lengths from the MFM images, averaging over multiple images for each lattice spacing. The observed exponential distribution is consistent with the strings being emergent thermal objects with weak mutual interactions. In the MFM images, they are frozen in place when the sample is cooled through the island moment blocking temperature. The greater fraction of longer strings for the larger lattice spacings is an expected result of a higher effective temperature relative to the interaction strength.

In order to better probe the apparent activated behavior seen in Figures 3e and 3f, we now consider the results of our XMCD-PEEM measurements. These data are taken on much thinner islands that can thermally reverse moment orientations on the time scale of the imaging. Figure 4a shows the average string length $\langle L \rangle$ taken from XMCD-PEEM data as a function of temperature for three different realizations of SFI with different lattice spacings. At low temperature, the average length is only weakly temperature-dependent, suggesting that the system is trapped in a metastable state due to proximity to the superparamagnetic blocking temperature (see Supplementary Note 4 for more details regarding the moment configuration). In all three cases, the average string length increases substantially at the highest temperatures, indicating the onset of more extensive thermal fluctuations among the moments.

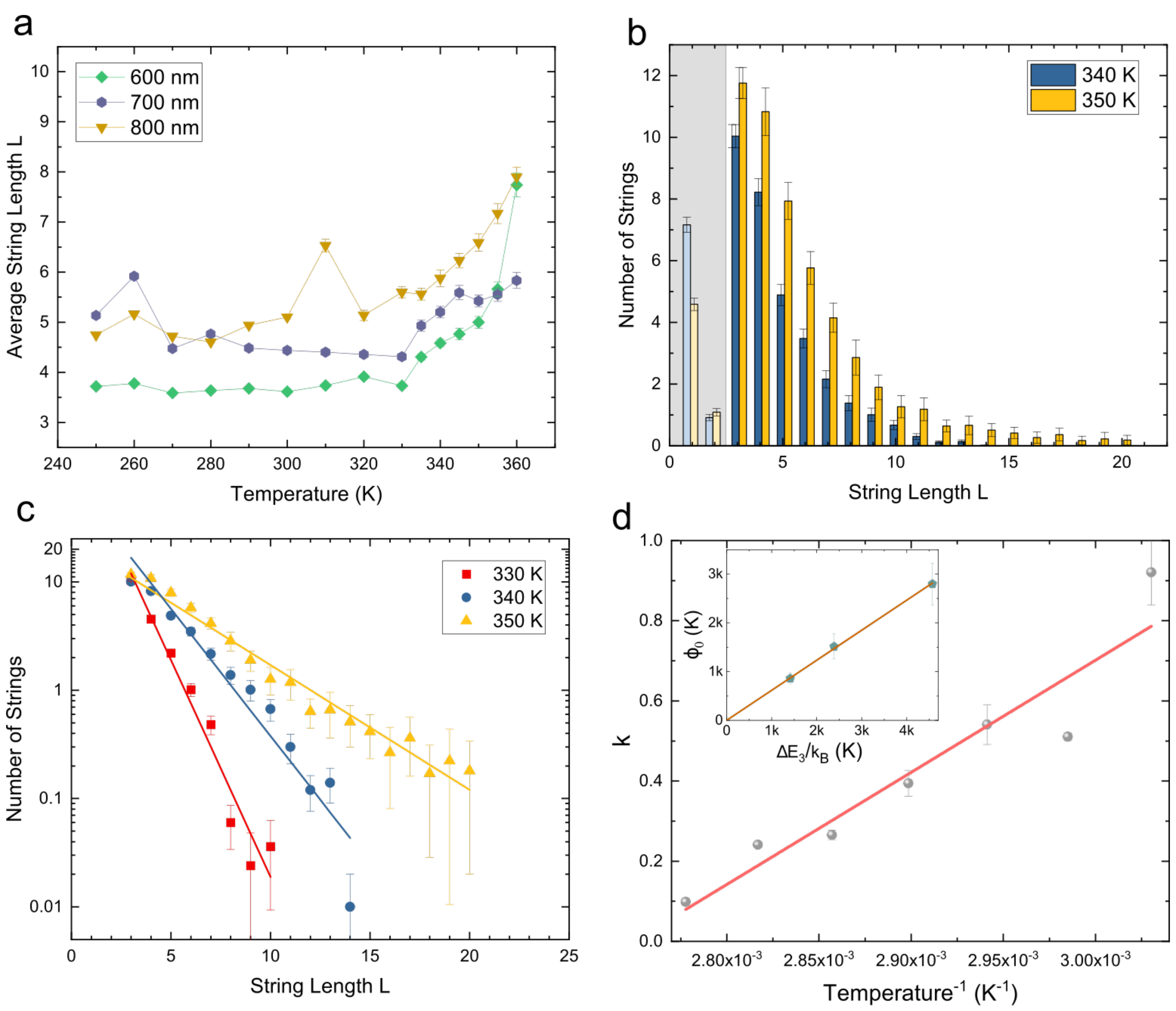

**Figure 4: String statistics from XMCD-PEEM data. (a)** The average length of strings as a function of temperature from the XMCD-PEEM data (measured in numbers of unhappy vertices). Note that the distribution is flat below $T \approx 330$ K, indicating that the system falls out of thermal equilibrium. **(b)** Distribution of strings with various length for the $a$ = 600 nm lattice, averaging over approximately 80 images (see Methods section). The grey section covers those strings that are connecting only contiguous interior plaquettes and therefore are expected to have a different energy scale. **(c)** The distribution of string length with a fit to a Boltzmann distribution $p(L) \propto e^{-Lk}$ as described in the text. The error bars in panels a-c represent the standard errors of the data taken at each temperature **(d)** The fitted $k$ from the Boltzmann distribution as a function of inverse temperature for $a$ = 600 nm and 330 K ≤ $T$ ≤ 360 K shows a linear dependence on the inverse temperature, $k(T) = \phi_0/T - \phi_1$. Inset: the fit parameter $\phi_0$ as a function of $[\Delta E_3/k_B]$ computed from micromagnetics for all three samples, where the error bars are the standard errors of the fit parameters.

Because the energy of a string is proportional to its length, one expects from the Boltzmann distribution that the probability $p(L)$ of finding a string of length $L$ is exponential when the system is thermalized, as seen for the high temperature annealing data discussed above. In other words, we expect that $p(L) \propto e^{-Lk}$ when the moments are thermalized, where $k$ is the reciprocal of a characteristic length and depends on temperature. Following a Boltzmann distribution, this corresponds to the average energy per unit length of the string, $\phi$, divided by temperature, or $k = \phi/T$.

We can relate these parameters to the microscopic behavior of the island moments. As strings at lower temperatures mostly connect $z = 3$ vertices, $\phi$ should be close to the energy difference between a Type I and a Type II vertex for $z = 3$, i.e., $\phi \approx \Delta E_3/k_B = (E(II_3) - E(I_3))/k_B$. Since $\phi$ depends on the magnetization of the moments, and we know from previous measurements that the magnetization of thin permalloy films is temperature dependent in this regime[29], $\phi$ must also be temperature dependent.

Figure 4b and 4c show the experimentally measured average distribution of string lengths, which is proportional to $p(L)$, confirming the expected exponential Boltzmann behavior described above. To find that dependence, we extract the reciprocal characteristic length $k$ as the slope of the linear fits in Figure 4c for experiments at different temperatures. In Figure 4d, we plot $k(T)$ vs. $1/T$, and we find a linear dependence, $k(T) = \phi_0/T - \phi_1$. This is equivalent to the energy cost per length, $\phi(T)$, having a linear temperature dependence of $\phi(T) = \phi_0 - \phi_1 T$. While the linear functional dependence of $\phi(T)$ cannot be easily justified over a wide temperature interval, a linear expression in our limited temperature range of measurement is a reasonable first order approximation.

Since we should be relatively close to the Curie temperature of permalloy for such thin islands[29], we can validate our model by estimating the Curie temperature from $\phi(T)$. Because the magnetization goes to zero at the Curie temperature, then $\phi(T)$ must also, and we can therefore take $\phi(T_{Curie}) = 0$, yielding $T_{Curie} \sim \phi_0 / \phi_1$. We obtain from this estimate that $T_{Curie}$ is of order 400 K for all three of our lattice spacings (see Supplementary Table 3), quite consistent with experimental results for very thin permalloy films[29].

We can further validate our fits from the magnitude of $\phi_0$. At low temperature, strings are composed primarily of $z = 3$ unhappy vertices, therefore $\phi_0$ must be close in magnitude to $\Delta E_3/k_B$, which can be computed from micromagnetics. In the inset of Figure 4d, we plot the fitted $\phi_0$ vs. $\Delta E_3/k_B$, determined from a micromagnetic calculation (see Supplementary Table 3). We see that the quantities are proportional, and that $(\Delta E_3/k_B)/\phi_0 \sim 1.60$. The deviation of the ratio from 1.0 is potentially attributable to limitations in the micromagnetic modelling (which, for example, does not include temperature dependence of the magnetization and interactions beyond the nearest neighbors), and to the fact that other types of vertex excitations also contribute to the string energy. The agreement is nonetheless striking in the sense that $\phi_0$ and $\phi_1$ are obtained not by fitting the data directly, but by fitting the temperature dependence of the Boltzmann parameter, $k$, which is itself derived from fits to data at each temperature. The results demonstrate that the physics of the topologically complex structure of SFI, with its large unit cell, can be robustly represented through the simple language of one-dimensional strings and their Boltzmann statistics.

## Discussion

The observed string correlations among unhappy vertices at low temperatures in SFI contrast sharply with the correlations among unhappy vertices in other vertex-frustrated systems. The correlations in the disordered portions of tetris ice are weak (i.e., exponentially decaying)[7,17] while Shakti ice has algebraic correlations[6,8,17,30], pointing to the criticality of its topological phase. In this case, the unhappy vertices in SFI are strongly correlated, forming one-dimensional collective emergent objects whose structure defines the magnetic moment ensemble, and whose motion and evolution define the overall kinetics of the system.

This string picture also has consequences for understanding the kinetics of the SFI system. As one-dimensional objects with ends fixed at the interior plaquettes, we can choose to consider the strings as defining a natural partition of the phase space into topological sectors. These sectors are then homotopy classes of the string configurations, i.e., ensembles of magnetic configurations corresponding to strings that are topologically equivalent, in that the strings can be continuously deformed without changing their ends. Therefore, topologically trivial kinetics correspond to bending and stretching modes of the strings that do not change the ends of the strings and thus do not alter the homotopy class. Relaxation and equilibration below the minimal energy of a sector requires the system to execute topologically non-trivial kinetics that can take it across homotopy classes. Such changes cut across homotopy classes, in the form of so-called string reconnection[20,22], constituting a physical manifestation of one-dimensional "topological surgery" in which strings cross, disconnect[31,32], and reconnect[33,34]. Similar phenomena are known theoretically to affect kinetics in some classical systems,[35,36] which evolve through rupture of the topological protection; they are likely to impact the kinetics here.

String correlations and their use in the partition of phase space into topological sectors are familiar in quantum topological matter, where they are a consequence of quantum entanglement[,23,24,33,34]. Our results demonstrate that even in a classical system, frustration can impose correlations on a disordered state that are strong enough to generate a string phase. The robust string physics of SFI demonstrates that artificial spin ice systems provide a platform for such a framework in a purely classical context. Future examination of the string kinetics in SFI and related artificial spin ice systems should enable exploration of the physics of string excitations and their topological properties in a well-characterized and easily controlled experimental system.

## Methods

### *Sample Preparation*

The artificial spin ice samples used in this work were fabricated through a process similar to that described in previous papers[6,7,8,]. We first used electron-beam lithography to write patterns on $Si/SiO_x$ substrates with spin-coated bilayer resists. Various thicknesses of permalloy ($Ni_{80}Fe_{20}$) films were then deposited into the patterns via ultrahigh vacuum electron beam evaporation, followed by aluminum capping layers of 2 nm thick for XMCD-PEEM samples and 3 nm thick for MFM samples to mitigate oxidation of the underlying permalloy.

### *Thermal Annealing and MFM Measurement*

The samples used for thermal annealing and MFM measurements had lateral island dimensions of 220 nm × 80 nm and thickness of about 15 nm. These dimensions were chosen so that the magnetic moments of the nanoislands were frozen at room temperature. Five SFI arrays were designed with lattice constants of 300 nm, 320 nm, 360 nm, 440 nm, and 480 nm. All arrays were polarized along the [1,1] direction then heated to 818 K at a rate of 10 K/min. The samples were then held at 818 K for 15 minutes before being cooled to 673 K at a rate of 1 K/min and then cooled to room temperature for measurement[29].

Two MFM scans were performed on each array to minimize location variances. Each MFM image contained about 900 total vertices and 1300 moments. Square ice lattices that were annealed simultaneously showed large ground state domains in MFM scans, demonstrating that the annealing protocols could successfully set vertices to the low energy states.

*XMCD-PEEM Measurement*

We conducted XMCD-PEEM experiments at the PEEM-3 endstation at beamline 11.0.1.1 of the Advanced Light Source, Lawrence Berkeley National Lab. The samples used for the XMCD-PEEM experiments had lattice constants of 600 nm, 700 nm, and 800 nm, and islands with approximate lateral dimensions of 470 nm × 170 nm. The exact dimensions measured by SEM are given in Supplementary Table 2. The permalloy thicknesses of PEEM samples were approximately 2.5 nm, and the island moments were thermally active in the tested temperature window. We heated the sample from 250 K to 360 K in 5-10 K steps and took 100 PEEM images at the Fe $L_3$ absorption edge at each temperature point. The 100 PEEM images consisted of ten exposures with a left-circularly polarized X-ray beam followed by ten exposures with a right-circularly polarized beam, repeated five times. Charging problems during the experiment caused a small fraction (≤20%) of the images to be out of focus, and these images were excluded from the analysis. The exposure time was set to 0.5 seconds and the total acquisition time at each temperature was about 150 seconds including computer read-out between exposures. The PEEM image field of view was set at 15 x 15 to 18 x 18 $\mu m^2$ and there were about 500 islands within each image. We use code prepared with MATLAB to extract the intensity for each island from every PEEM image. The intensity values usually fall into two groups, one with higher average and the other with lower average. We can then resolve the moment direction for each single island for each image. The island flip rate was obtained by calculating the fraction of islands that changed their moment directions between two sequential PEEM images with the same x-ray polarity taken at each temperature, divided by the acquisition time of two images. In addition to the data shown in Figure 4, additional PEEM measurements and detailed associated parameters are discussed in Supplementary Note 4.

**AUTHORS' CONTRIBUTIONS**

J. Ramberger, J. Batley, and J. D. Watts performed film depositions under the guidance of C. Leighton, and D. Bromley prepared other samples under the guidance of L. O'Brien, with X. Zhang, Y. Lao, J. Sklenar, and N. S. Bingham overseeing the lithography. X. Zhang, Y. Lao, J. Sklenar, N. S. Bingham, H. Saglam and R. V. Chopdekar performed the PEEM characterization of the thermally active samples. X. Zhang and N. S. Bingham performed above-Curie point annealing and MFM characterization. X. Zhang and H.

Saglam performed micromagnetic calculations, and X. Zhang and S. Subzwari analyzed the string structures. A. Duzgun performed Monte Carlo and spin dynamics simulations, under the guidance of C. Nisoli. C. Nisoli provided the string picture of the ground state and kinetics and wrote the first draft. P. Schiffer supervised the entire project. All authors contributed to the discussion of results and to the finalization of the manuscript.

**FUNDING ACKNOWLEDGEMENT**

We are grateful to I-A. Chioar for helpful discussions. Work at Yale University and the University of Illinois at Urbana-Champaign was funded by the US Department of Energy, Office of Basic Energy Sciences, Materials Sciences and Engineering Division under Grant No. DE-SC0010778 and Grant No. DE-SC0020162. This research used resources of the Advanced Light Source, a DOE Office of Science User Facility under contract no. DE-AC02-05CH11231. Work at the University of Minnesota was supported by NSF through Grant No. DMR-1807124. Work at the University of Liverpool was supported by the UK Royal Society, Grant No. RGS\R2\180208. Work at Los Alamos National Laboratory was carried out under the auspices of the US Department of Energy through Los Alamos National Laboratory, operated by Triad National Security, LLC (Contract No. 892333218NCA000001) and financed by DoE LDRD.

**DATA AVAILABILITY**

Experimental and simulation data generated in this study have been deposited in the Dryad database under DOI:10.5061/dryad.jdfn2z3c2[37]

**COMPETING INTERESTS**

The authors declare no competing interests.

**Supplementary Information**

**String Phase in an Artificial Spin Ice**

Xiaoyu Zhang[1,2,3], Ayhan Duzgun[4], Yuyang Lao[2,3], Shayaan Subzwari[1], Nicholas S. Bingham[1], Joseph Sklenar[2,3,5], Hilal Saglam[1], Justin Ramberger[6], Joseph T. Batley[6], Justin D. Watts[6,7] Daniel Bromley[8], Rajesh V. Chopdekar[9], Liam O'Brien[8], Chris Leighton[6], Cristiano Nisoli[4], & Peter Schiffer[1,2,3,10]

[1]*Department of Applied Physics, Yale University, New Haven, CT 06511, USA*

[2]*Department of Physics, University of Illinois at Urbana-Champaign, Urbana, Illinois 61801, USA*

[3]*Frederick Seitz Materials Research Laboratory, University of Illinois at Urbana-Champaign, Urbana, Illinois 61801, USA*

[4]*Theoretical Division and Center for Nonlinear Studies, MS B258, Los Alamos National Laboratory, Los Alamos, New Mexico 87545, USA*

[5]*Department of Physics and Astronomy, Wayne State University, Detroit, Michigan 48201, USA*

[6]*Department of Chemical Engineering and Materials Science, University of Minnesota, Minneapolis, Minnesota 55455, USA*

[7]*School of Physics and Astronomy, University of Minnesota, Minneapolis, Minnesota 55455, USA*

[8]*Department of Physics, University of Liverpool, Liverpool L69 3BX, United Kingdom*

[9]*Advanced Light Source, Lawrence Berkeley National Laboratory, Berkeley, CA 94720, USA*

[10]*Department of Physics, Yale University, New Haven, CT 06511, USA*

**Supplementary Note 1: Monte Carlo simulations**

In order to better understand the Santa Fe Ice system, we have run extensive Monte Carlo simulations. We considered a system with 4 x 4 unit cells of the lattice (the unit cell is shown in Figure 1a in the main text), which consists of 8 x 8 composite cells or 1536 spins. We impose periodic boundary conditions and calculate the system energy using a vertex model where only nearest-neighbor interactions between collinear and perpendicular islands are included (*Supplementary Fig. 1*). The two spin interaction energies, $\epsilon_{\perp}$ and $\epsilon_{\parallel}$ correspond to the coupling between moments that are converging in the vertex perpendicular and parallel to each other respectively and are defined in *Supplementary Fig. 1*. In other words, in this approximation, the spins only interact with other spins that share a common vertex. The value of $\epsilon_{\parallel}$ defines an energy (and thus temperature) scale which naturally depends on the details of the island geometry and spacing in a real system.

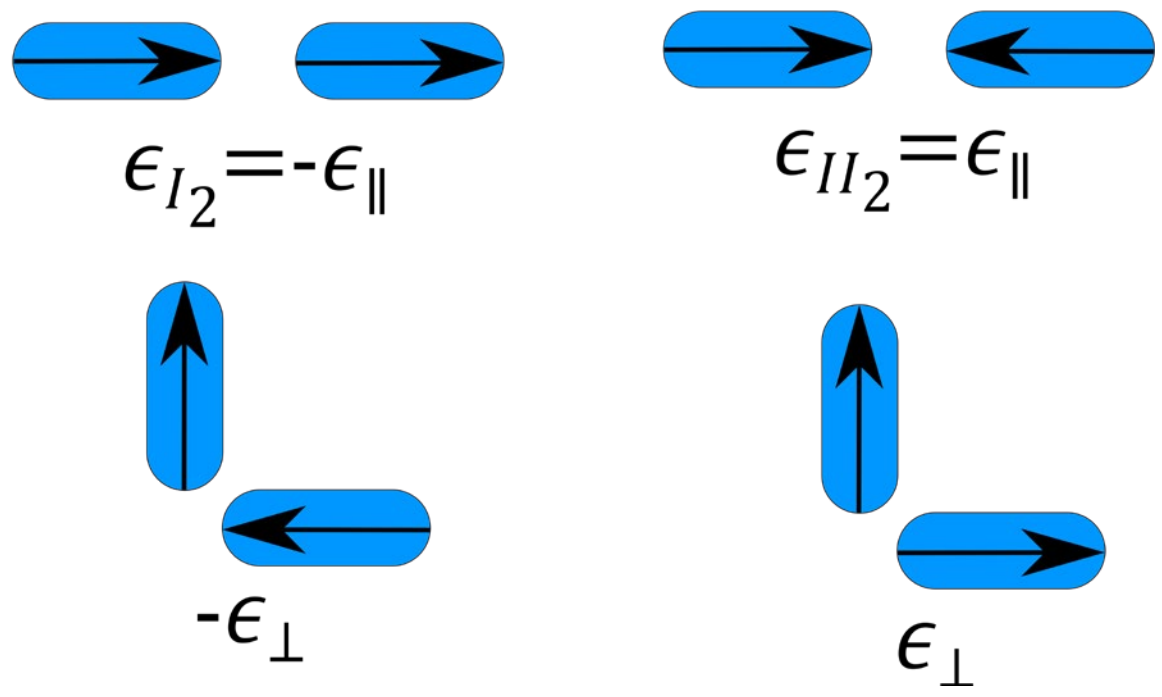

*Supplementary Fig. 1. Schematics defining the couplings in the Metropolis Monte Carlo simulations.*

The value of $\frac{\epsilon_\perp}{\epsilon_\parallel}$ determines the ground state of the system as shown in *Supplementary Fig. 2*. The two possible ground states are either a long-range-ordered state or a disordered string state with strings all of length three excitations. Note that, even in the case of a long-range antiferromagnetic ground state, the excitation profile is made of strings, due to the topological constraints for the number of unhappy vertices that can connect to a given plaquette. In *Supplementary Table 1*, we show the predicted population statistics of vertices at high temperature and zero temperature. Values at high temperature are merely the relative counts of the vertex type for a random distribution of moments. Values at zero temperature are different for the two different possible ground states. For the disordered string ground state, vertices of coordination two and four are all in their lowest energy configuration (Type $I_2$ and Type $I_4$), whereas 3/14 of the z = 3 vertices in a composite cell are unhappy vertices, i.e., in the Type $II_3$ moment configuration. For the long-range-ordered state, vertices of coordination $z$ = 3 and $z$ = 4 are all in their lowest energy configuration (Type $I_3$ and Type $I_4$), whereas all of the $z$ = 2 vertices are unhappy vertices, i.e., in the Type $II_2$ moment configuration. As a full long-range-ordered state was not seen in the measurements reported in this work, we limit ourselves to simulations for ratios below the threshold that would lead to long range order.

| Vertex type | High temperature state | Disordered string ground state vertex fractions | Long-range-ordered ground state vertex fractions |
|---|---|---|---|
| $I_4$ | ≈ 2/16 = 0.125 | 1 | 1 |
| $II_4$ | ≈ 4/16 = 0.25 | 0 | 0 |
| $III_4$ | ≈ 8/16 = 0.5 | 0 | 0 |
| $IV_4$ | ≈ 2/16 = 0.125 | 0 | 0 |
| $I_3$ | ≈ 2/8 = 0.25 | 11/14 ≈ 0.79 | 1 |
| $II_3$ | ≈ 4/8 = 0.5 | 3/14 ≈ 0.21 | 0 |
| $III_3$ | ≈ 2/8 = 0.25 | 0 | 0 |
| $I_2$ | ≈ 2/4 = 0.5 | 1 | 0 |
| $II_2$ | ≈ 2/4 = 0.5 | 0 | 1 |

*Supplementary Table 1. Predicted vertex fractions at high temperature and in the two ground states (i.e., at zero temperature).*

For the simulations, the system was cooled through 500 steps of exponentially decaying temperature values which approximately corresponds to setting $T(i+1) = 0.99\ T(i)$ for $i^{th}$ and $i+1^{th}$ steps. At each temperature step, 300,000 trials per spin were performed. The system energy was recorded after each of the last 60% of these trials (180,000 per spin), and these were used to calculate the heat capacity. The average of the vertex counts over the duration of all of the 300,000 updates at each temperature was used to calculate the relative populations.

The simulation procedure is as follows. We start with very high temperature and slowly anneal to zero temperature using a standard Metropolis algorithm, which uses single spin flips that mimic the thermal evolution seen in the PEEM experiments. We collect statistics of vertex populations, and we compute the heat capacity and the entropy during the annealing process. The heat capacity $C_v$ per spin is calculated from thermal fluctuations using the equation

$$\frac{C_v}{k_B} = \frac{1}{(k_B T)^2} \frac{\sigma^2(E_N)}{N} \tag{1}$$

where $E_N$ is the total energy of the system which has $N$ spins, and $\sigma$ is the standard deviation of $E_N$. The entropy per spin is calculated by

$$\frac{S(T)}{k_B} = \ln 2 + \int_{\infty}^{T} \frac{C_v}{k_B} \frac{dT}{T} \ . \tag{2}$$

*Supplementary Fig. 3 – Fig. 5* show snapshots for $\frac{\epsilon_\perp}{\epsilon_\parallel} = 1.3$. For these interactions, within the vertex approximation, the ground state corresponds to the disordered string ground state. The figures show a relatively high temperature state, $k_B T/\epsilon_\parallel = 1.35$, an intermediate temperature state, $k_B T/\epsilon_\parallel = 0.40$, and a configuration near $T = 0$. As the system is cooled, the high density of strings reduces to the disordered string ground state, where strings start and end at nearest interior plaquettes. At sufficiently low temperature, essentially all of the $z = 4$ vertices are of Type $I_4$ due to its low energy.

*Supplementary Fig. 6* shows the corresponding vertex statistics of $z = 4$, $z = 3$, and $z = 2$ vertex types and the curve for entropy as the system is cooled. In the high temperature limit, populations are entropic, i.e., proportional to their relative multiplicity, as also listed in *Supplementary Table 1*. Note the two bumps in the specific heat. The one at larger temperature denotes the crossover into a regime where the vertices obey the ice rule. The one at lower temperature denotes the crossover to the disordered string ground state. Note also the residual entropy at zero temperature demonstrating the extensive degeneracy of the ground state. Finally, note that the population of unhappy vertices, or Type $II_3$, do not tend to zero as temperature is lowered, because they form the strings. Naturally, as for the case of pyrochlore ice and honeycomb ice, the dipolar interaction, here neglected, might induce ordering at very low temperature.

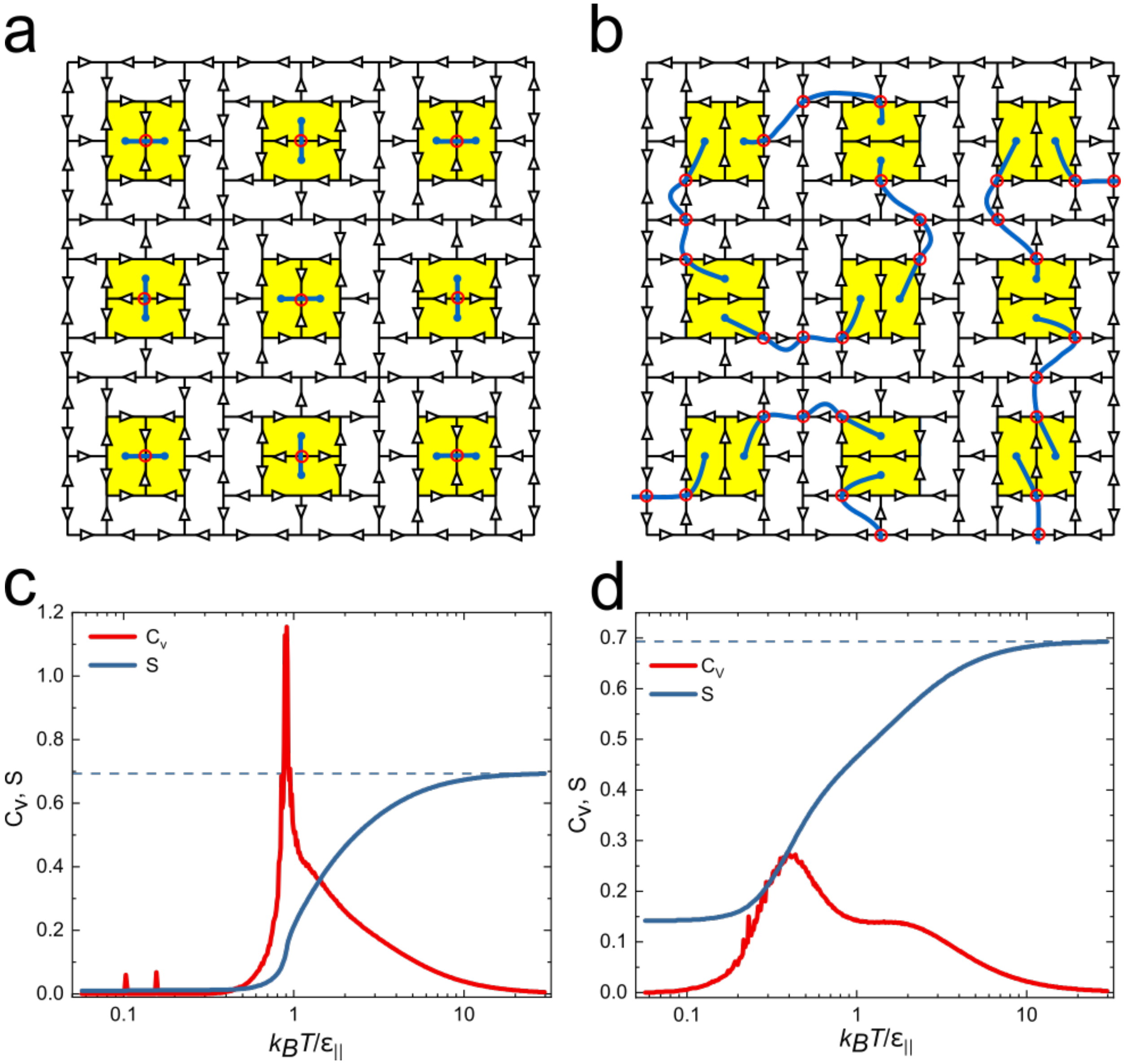

*Supplementary Fig. 2: Santa Fe ice ground states. Schematic of the moment configurations and the strings, showing (a) the long-range-ordered ground state, and (b) the disordered string ground state. (c) and (d) specific heat ($C_V$) and entropy (S) per moment as a function of effective temperature from Monte Carlo calculations for $\frac{\epsilon_\perp}{\epsilon_\parallel}$= 1.9 and 1.3, respectively. The simulations yield the long-range-ordered state and the disordered string ground state in these two cases. Note the sharp peak associated with the transition to long range order, the broad peak associated with the evolution to the disordered ground state, and the shoulder in each case associated with the development of short-range order on the vertices (the small low-temperature peaks are noise in the simulation). Note the residual entropy in the entropy curve for the disordered state. Entropy at infinite temperature is ln2 (dashed blue line).*

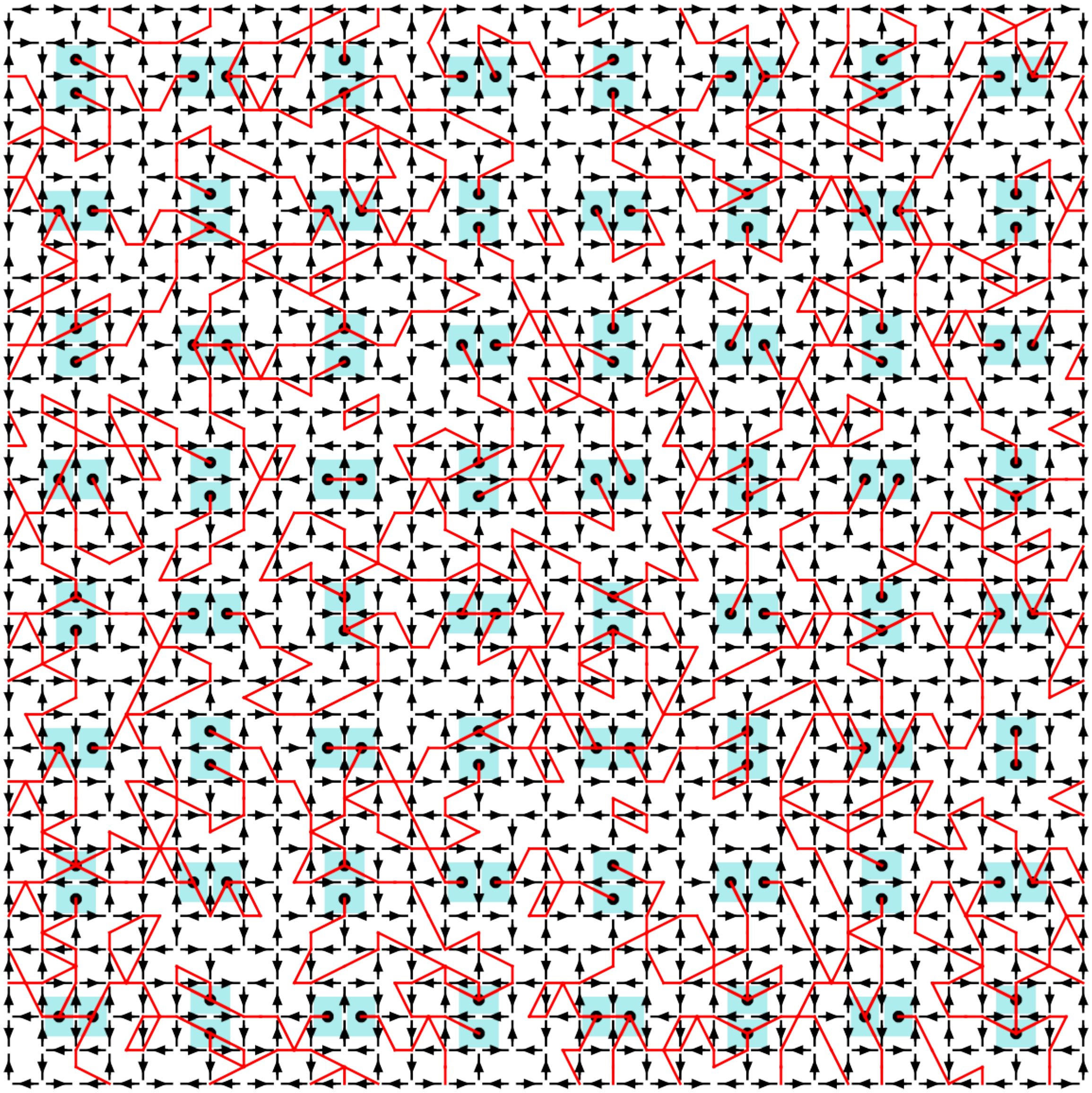

*Supplementary Fig. 3. Snapshots from Monte Carlo simulation for $k_B T/\epsilon_{\parallel}$ =1.35 and $\frac{\epsilon_{\perp}}{\epsilon_{\parallel}}$= 1.3. Red lines show the strings of unhappy vertices, and the full system size of 8 × 8 composite cells is shown. Note that, at this relatively high temperature, the high density of excitations leads to a higher density of strings and the presence of longer strings and loops.*

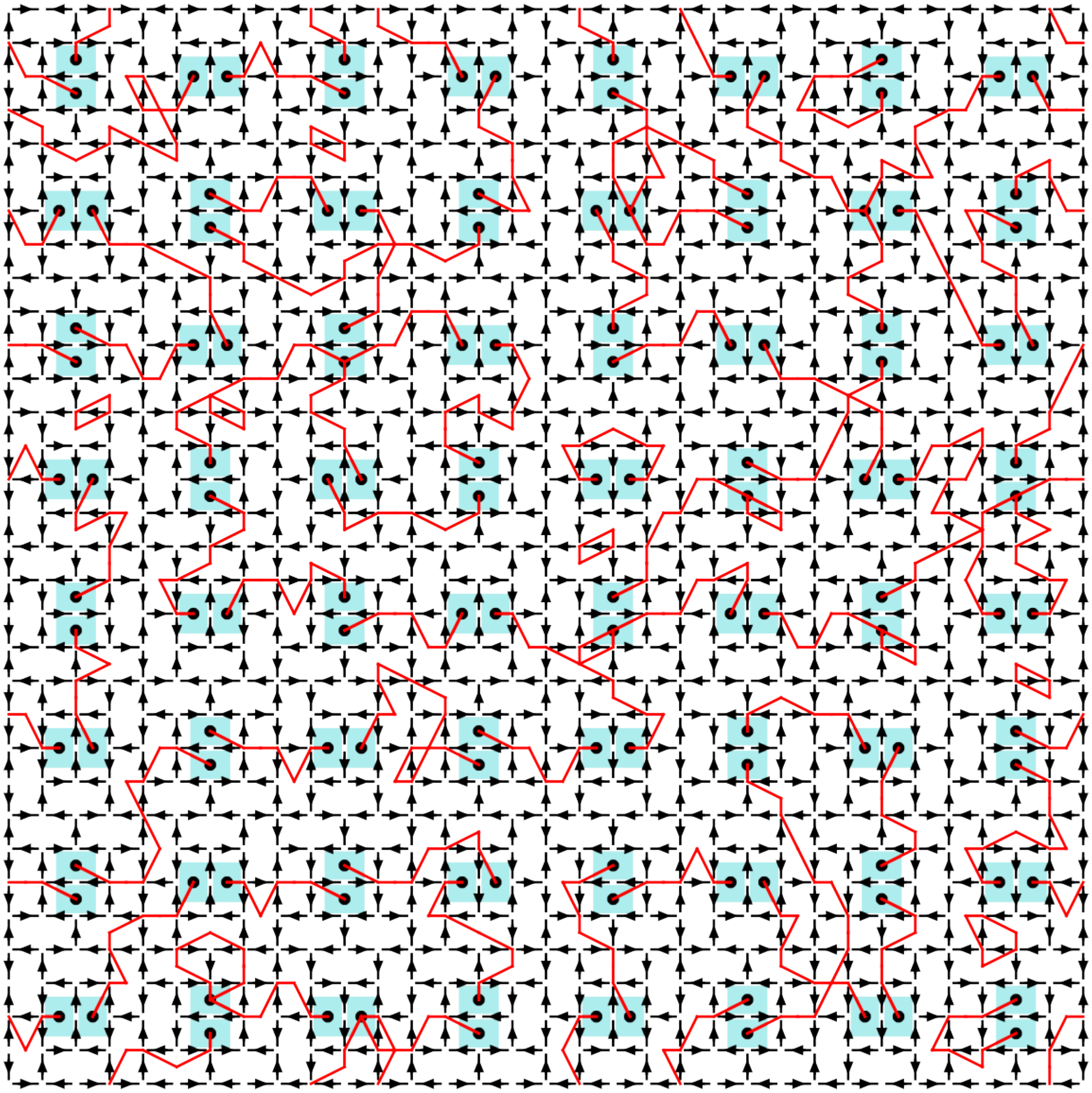

*Supplementary Fig. 4. Snapshots from Monte Carlo simulation for* $k_B T/\epsilon_\parallel$*= 0.40 and* $\frac{\epsilon_\perp}{\epsilon_\parallel}$*= 1.3. Red lines show the strings of unhappy vertices, and the full system size of 8 × 8 composite cells is shown. The picture shows an excited string state where strings are generally longer than the three unhappy vertices predicted in the ground state and can run between interior plaquettes farther away.*

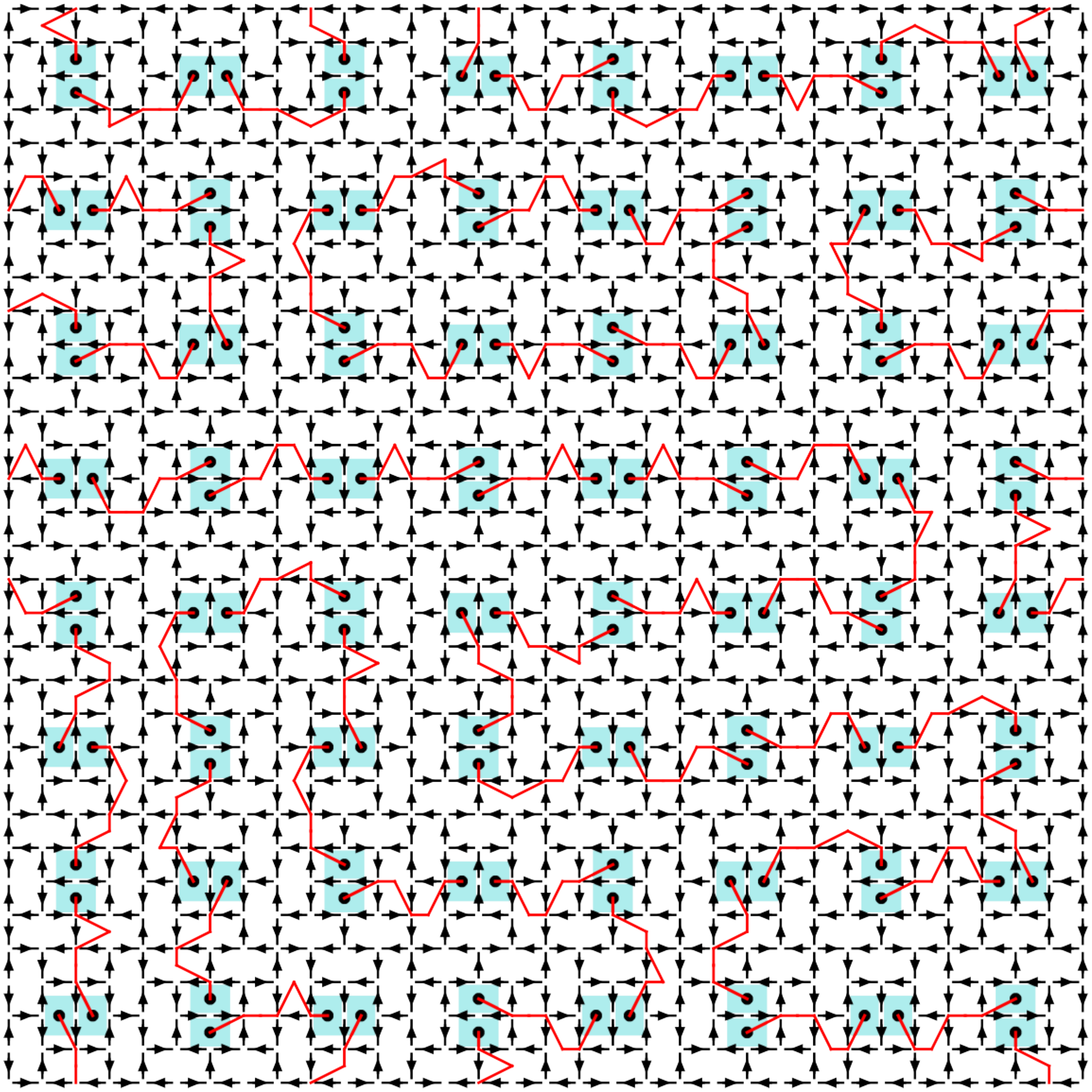

*Supplementary Fig. 5. Snapshots from Monte Carlo simulation for $k_B T/\epsilon_{\parallel}$ = 0.05 and $\frac{\epsilon_{\perp}}{\epsilon_{\parallel}}$= 1.3. Red lines show the strings of unhappy vertices, and the full system size of 8 × 8 composite cells is shown. The system here is in the disordered string ground state.*

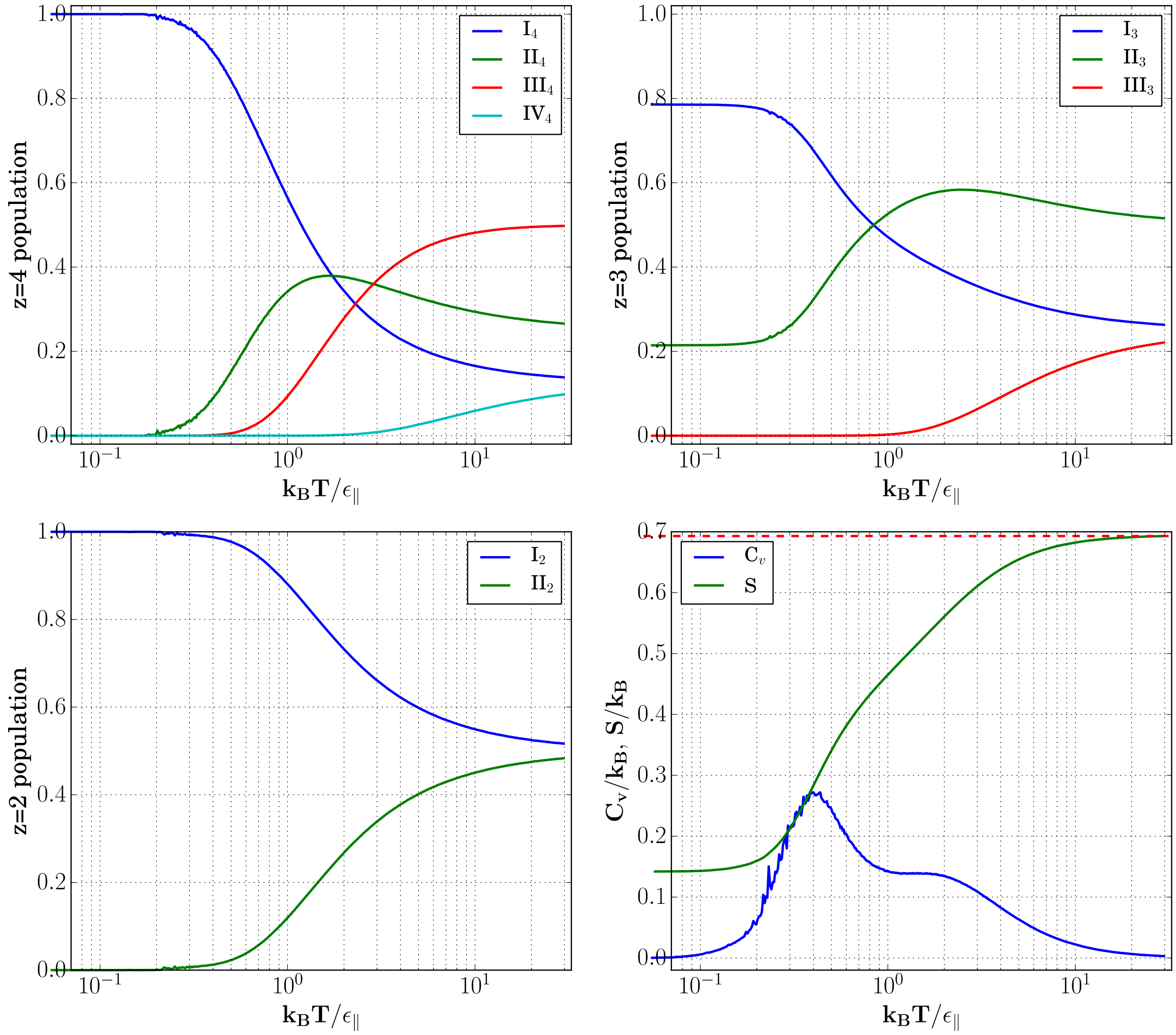

*Supplementary Fig. 6. Simulated vertex statistics for $\frac{\epsilon_\perp}{\epsilon_\parallel}$= 1.3. Relative counts of vertices for coordination numbers z = 4, z = 3, z = 2 as well as entropy and specific heat are shown (same as for Supplementary Fig. 2, included here for easy comparison). Note that all vertex types tend to their ground states except z=3 vertices: the unhappy vertices, or Type $II_3$, do not tend to zero as temperature is lowered, because they form the strings. The two bumps in the specific heat signal onset of the ice-rule, and then of the string state. Note also the residual entropy at T=0. Dashed red line in the entropy graph indicates ln 2.*

**Supplementary Note 2: Peripheral and Interior Plaquettes: Numbers of Connected Unhappy Vertices**

Santa Fe Ice has multiple coordination numbers at the different vertices, in that vertices can have two, three and four islands, or coordination numbers of $z$ = 2, 3, and 4. Within the SFI structure, the interior and peripheral plaquettes differ in that the interior plaquette has a $z$ = 2 vertex, while the peripheral plaquette vertices are $z$ = 3 or 4. This can be readily seen in *Supplementary Fig. 7*.

For the interior plaquettes, to enable each of the $z$ = 3 vertices to be "happy", i.e., in its vertex ground state, the $z$ = 2 vertex must be in its excited "unhappy" state of the two moments in opposite directions, as can be readily seen in *Supplementary Fig. 7 b*. Flipping one of the moments in the $z$ = 2 vertex to put it into its ground state would simply make one of the $z$ = 3 vertices "unhappy". The structure thus requires that the interior plaquettes be connected by at least one unhappy vertex around its edges, although it is easy to see that reversing other moments could produce any odd number of unhappy vertices, e.g., either one or three or five. Those plaquettes are therefore intrinsically frustrated in that they cannot have all vertices in a ground state, and their frustration affects nearby, unfrustrated peripheral plaquettes.

For the peripheral plaquettes, it is possible to put all of the vertices around the plaquette in their lowest energy "happy" state, as demonstrated in *Supplementary Fig.7 c* and d. By flipping any of the moments, one would create a pair of unhappy vertices on either side of that moment, and one could create either four or six unhappy vertices by strategic choice of which moments to flip. Therefore, the peripheral plaquettes must be connected by an even number of unhappy vertices (including possibly zero vertices as shown in the figure). These two conditions lead naturally to the string construction described in the main text

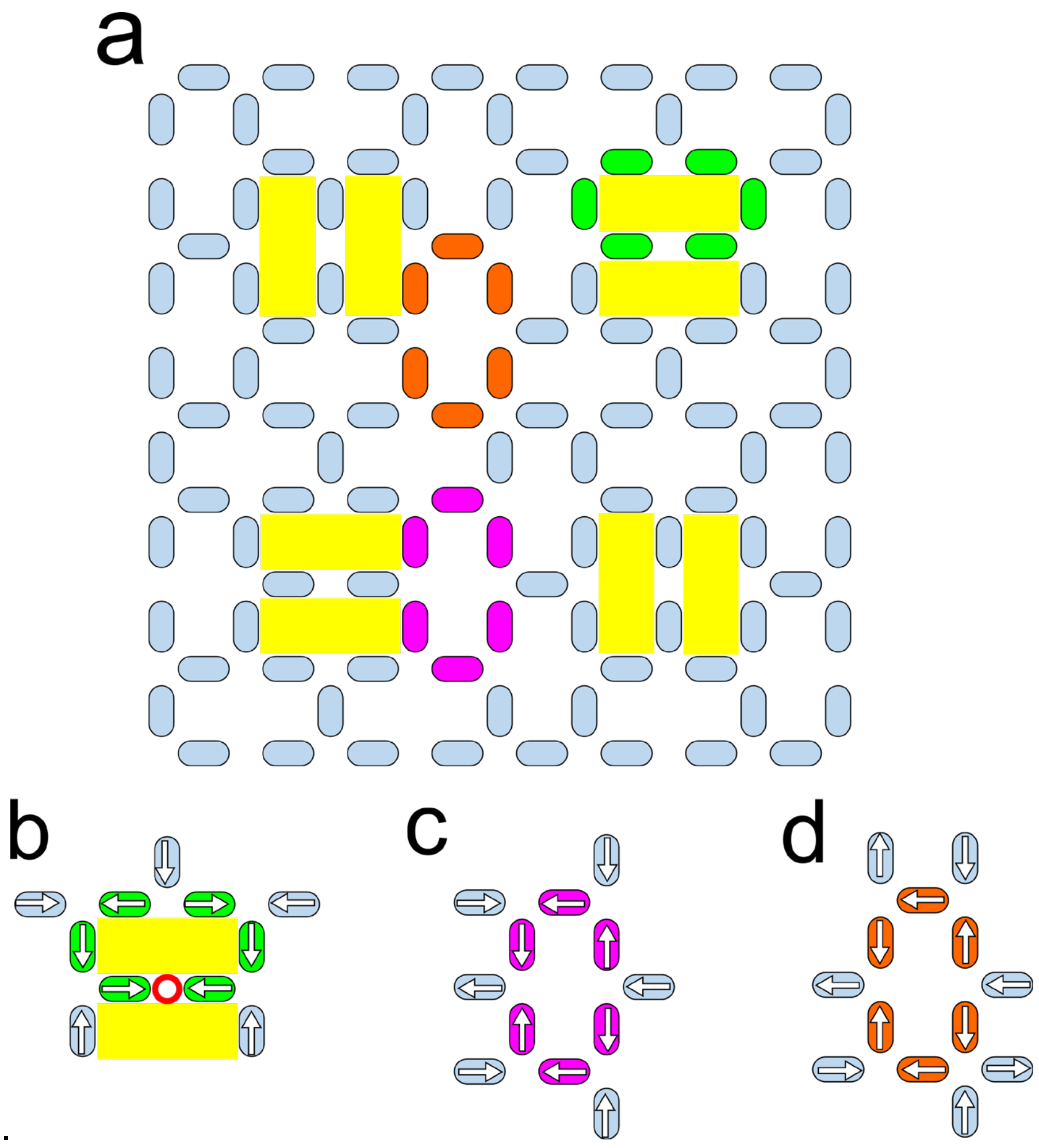

*Supplementary Fig. 7. Santa Fe Ice structure and sample plaquettes. (a) A full unit cell of the SFI structure with interior and peripheral plaquettes highlighted in green (interior) and pink and orange (peripheral). (b) The sample interior plaquette with moment directions indicated. There is a single unhappy vertex associated with the z = 2 vertex, emphasized with a red circle. Flipping a moment around the edges of the plaquette would simply place one of the z = 3 vertices into an unhappy state or add a pair of unhappy vertices – therefore the plaquette must have an odd number of unhappy vertices. (c,d) Sample peripheral plaquettes with moment directions indicated. The two plaquettes have all z = 3 vertices (for c), or a single z = 4 vertex (for d). In both cases the moments can be assigned as shown, so that all of the vertices are in their ground state, i.e., all are happy. Flipping any moment around the plaquette would result in a pair of unhappy vertices, and therefore these plaquettes must have an even number of unhappy vertices (including zero as a possible even number).*

**Supplementary Note 3: Additional MFM data**

Below we provide sample MFM images after high-temperature annealing (details of annealing provided in the Methods section). We repeated the annealing experiment two times on two different samples. To minimize the sample variance and location variance, we took two MFM images at different locations on each array after each annealing run and calculated the vertex fraction, shown in Figure 3 in the main text. *Supplementary Fig. 8* gives examples of MFM images of SFI arrays with different lattice constants.

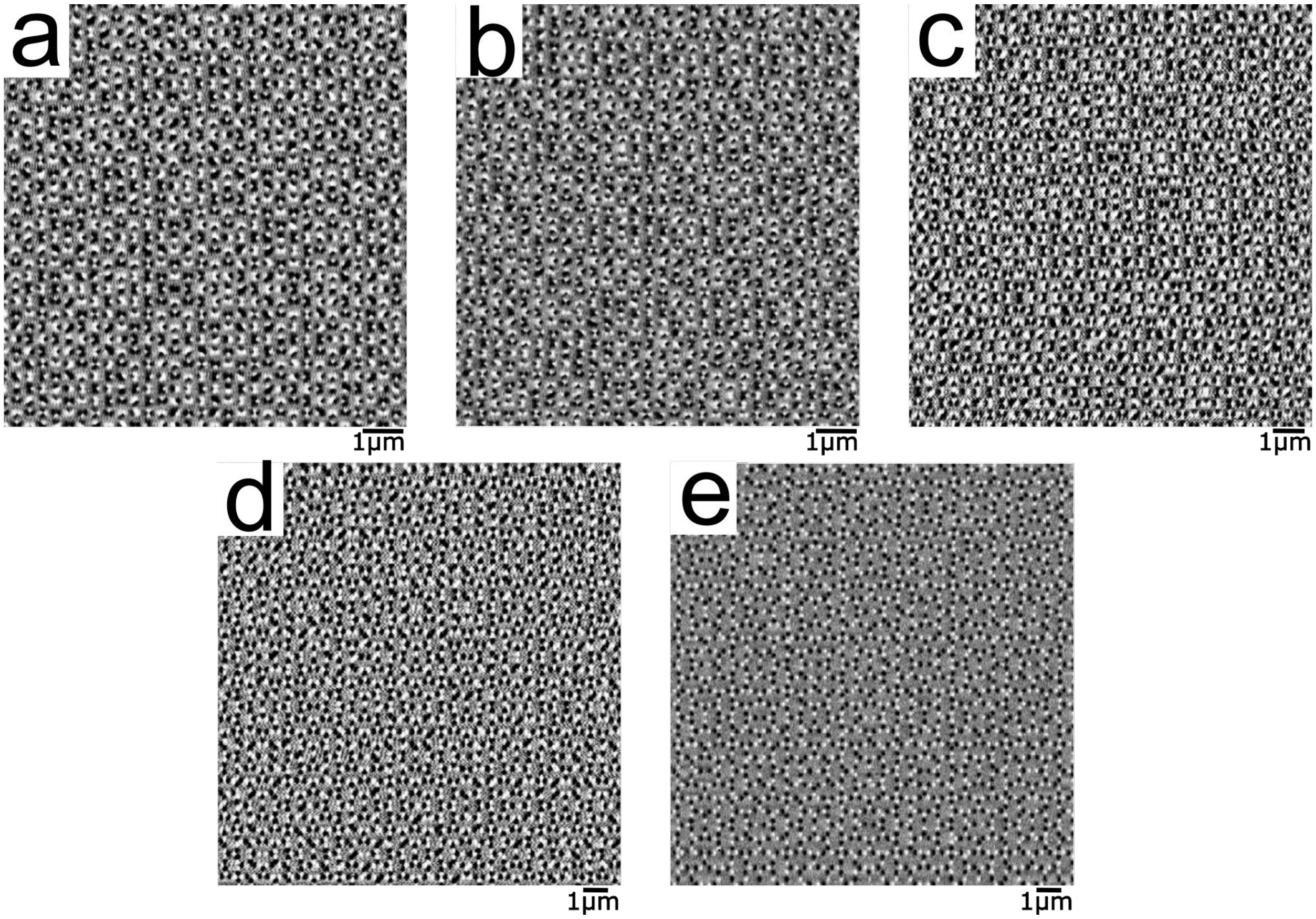

*Supplementary Fig. 8. MFM images of SFI lattice with lattice constant a) 300 nm, b) 320 nm, c) 360 nm, d) 440 nm, e) 480 nm after annealing at 818 K for 15 mins.*

**Supplementary Note 4: Additional PEEM data**

We performed five different runs of PEEM measurements on SFI, with different samples in each case. The exact island dimensions for each run are shown in *Supplementary Table 2* For the first, second, fourth, and fifth PEEM runs, we focused on dynamics, taking a series of 100 PEEM images at each temperature as samples were stepwise heated or cooled. The 100 PEEM images consisted of ten exposures at the Fe $L_3$ edge with a left-circularly polarized X-ray beam followed by ten exposures with a right-circularly polarized beam, repeated five times. The exposure time was set to 0.2 to 0.5 seconds and the total acquisition time at each temperature was about 130 to 150 seconds including computer read-out between exposures. The PEEM images were analyzed as described in the Methods section in the main text. For the third PEEM run, we took two PEEM-XMCD images at 10 different locations at each temperature point to obtain better statistics. Each image was constructed from four PEEM images with a left-polarized X-ray beam and four with a right-polarized beam; we took the average of every four PEEM images and subtracted the averaged images with left-polarized X-rays from the right-polarized X-rays to obtain a PEEM-XMCD image. The exposure time for each PEEM image was 0.7 seconds for better intensity statistics per image.

We did not observe significant temperature dependence in the moment configuration from the first four PEEM experiments. Presumably the moments were trapped in metastable states for those samples in our temperature range, due to the topological complexity of the lattice combined with structural disorder associated with the lithography. The detailed results for these experiments are therefore not included in this paper, but they are available upon request. The data in the remainder of this section and in the main text is from the fifth run, where significant temperature dependence was observed at the upper end of the temperature range measured, enabling us to study the thermal properties of the strings. The results in the remainder of this section focus on data from the fifth run.

| PEEM Experiments | 600 nm SF array | | 700 nm SF array | | 800 nm SF array | |
|---|---|---|---|---|---|---|
| | Length (nm) | Width(nm) | Length (nm) | Width(nm) | Length (nm) | Width(nm) |
| First | 430±3.9 | 143±6.2 | 455±2.8 | 160±2.6 | 484±4.3 | 175±2.2 |
| Second | 439±6.5 | 146±6.5 | 448±5.2 | 156±2.2 | 471±1.2 | 169±3.2 |
| Third | 455±5.2 | 173±2.1 | 484±4.5 | 192±2.4 | 498±1.7 | 193±1.6 |
| Fourth | 470±3.6 | 191±8.6 | 501±3.7 | 205±5.2 | 511±6.3 | 210±2.6 |
| Fifth | 482±1.5 | 189±1.5 | 477±4.9 | 186±3.8 | 477±1.2 | 179±1.5 |

*Supplementary Table 2. Island dimensions for all five PEEM runs.*

We calculated the vertex statistics and island flip rate as shown in *Supplementary Fig. 9*. *Supplementary Fig. 10* shows the plot of $k$ vs. $1/T$ with a linear fitting for (a) 700 nm and (b) 800nm. The $\phi_0$, the slope of $k$ vs. $1/T$, is compatible with $\Delta E_3/k_B$, where the exact numbers are given in *Supplementary Table 3*.

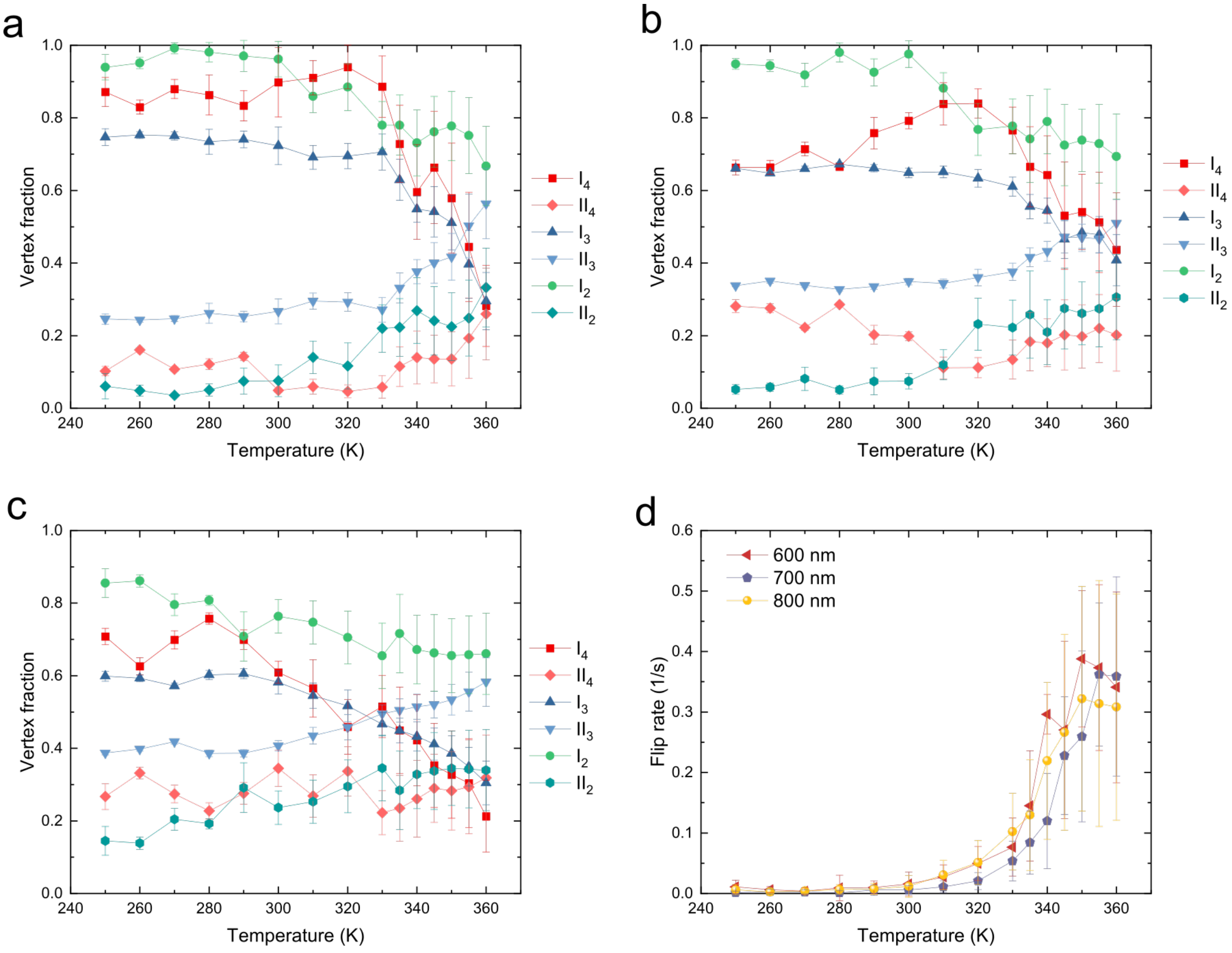

*Supplementary Fig. 9. Vertex statistics for Type I vertices and Type II vertices for vertex coordination of four, three, and two, colored in red, blue, and green, respectively for (a) 600 nm, (b) 700 nm, and (c) 800 nm SFI. (d) Average islands flip rate as a function of temperature for a= 600 nm (red), 700 nm (violet), and 800 nm (yellow) SFI. The error bars represent the standard deviations of the data.*

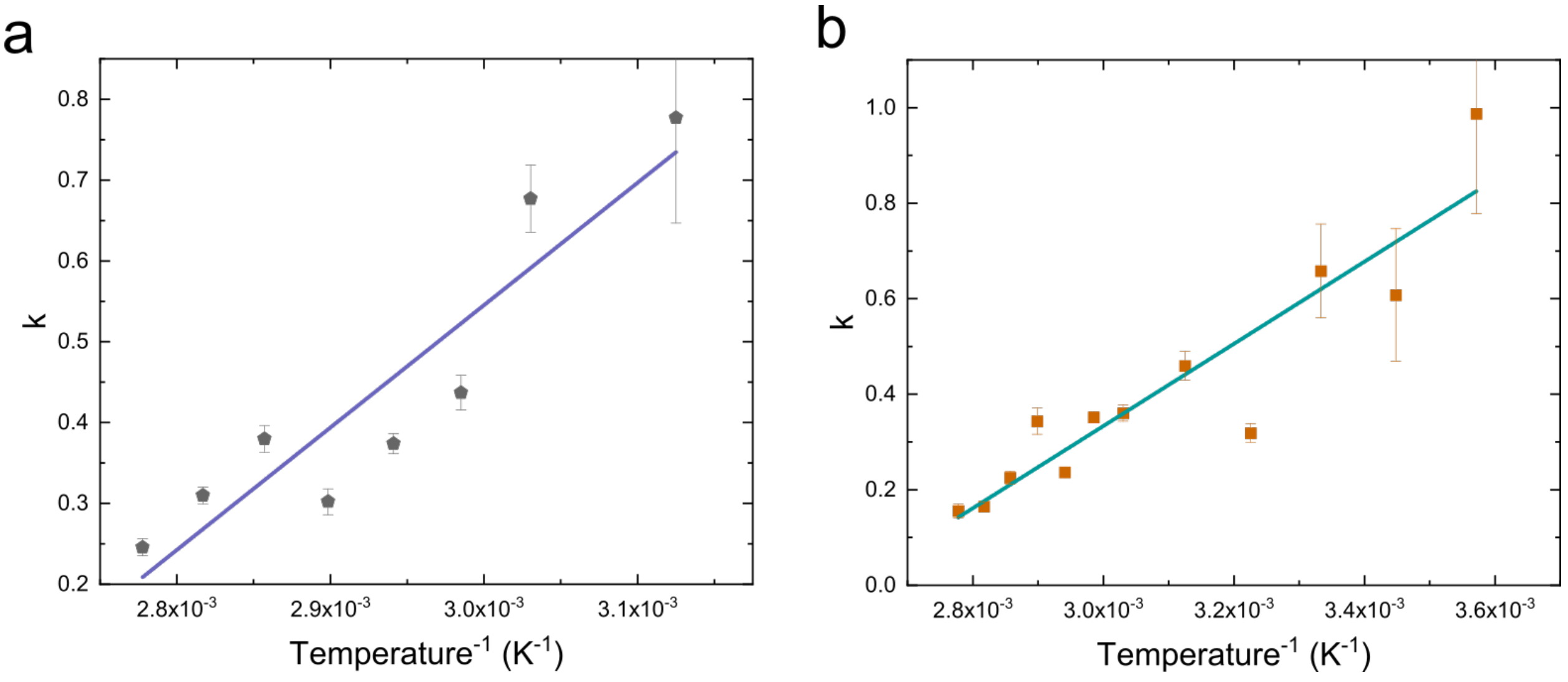

*Supplementary Fig. 10,* The string tension fit from the Boltzmann distribution as a function of inverse temperature for *(a) 700 nm for 320 K ≤ T ≤ 360 K and (b) 800 nm SFI for 280 K ≤ T ≤ 360.* The error bars are the standard errors of the fit parameters.

| SFI array | $E_{I2}$ (E-18 J) | $E_{II2}$ (E-18 J) | $E_{I3}$ (E-18 J) | $E_{II3}$ (E-18 J) | $\Delta E_3/k_B$ (1000 K) | $\phi_0$ (100 K) | $\phi_1$ | $\phi_0/\phi_1$ (K) |
|---|---|---|---|---|---|---|---|---|
| 600 nm | 1.42 | 1.55 | 2.00 | 2.06 | 4.57 | 27.98 | 7.69 | 364 |
| 700 nm | 1.41 | 1.47 | 2.07 | 2.10 | 2.39 | 15.14 | 4.00 | 378 |
| 800 nm | 1.39 | 1.42 | 2.05 | 2.07 | 1.41 | 8.60 | 2.25 | 382 |

*Supplementary Table 3. Magnetostatic energies for the ground state and the first excited state of Z = 2 and 3 vertices, calculated by micromagnetic simulation program MUMAX3 [1], using the island dimensions for the fifth PEEM run. Also shown are the fit parameters $\phi_0$ and $\phi_1$. As described in the text, $\phi_0$/ $\phi_1$ gives an approximate value of the Curie temperature of the permalloy films.*

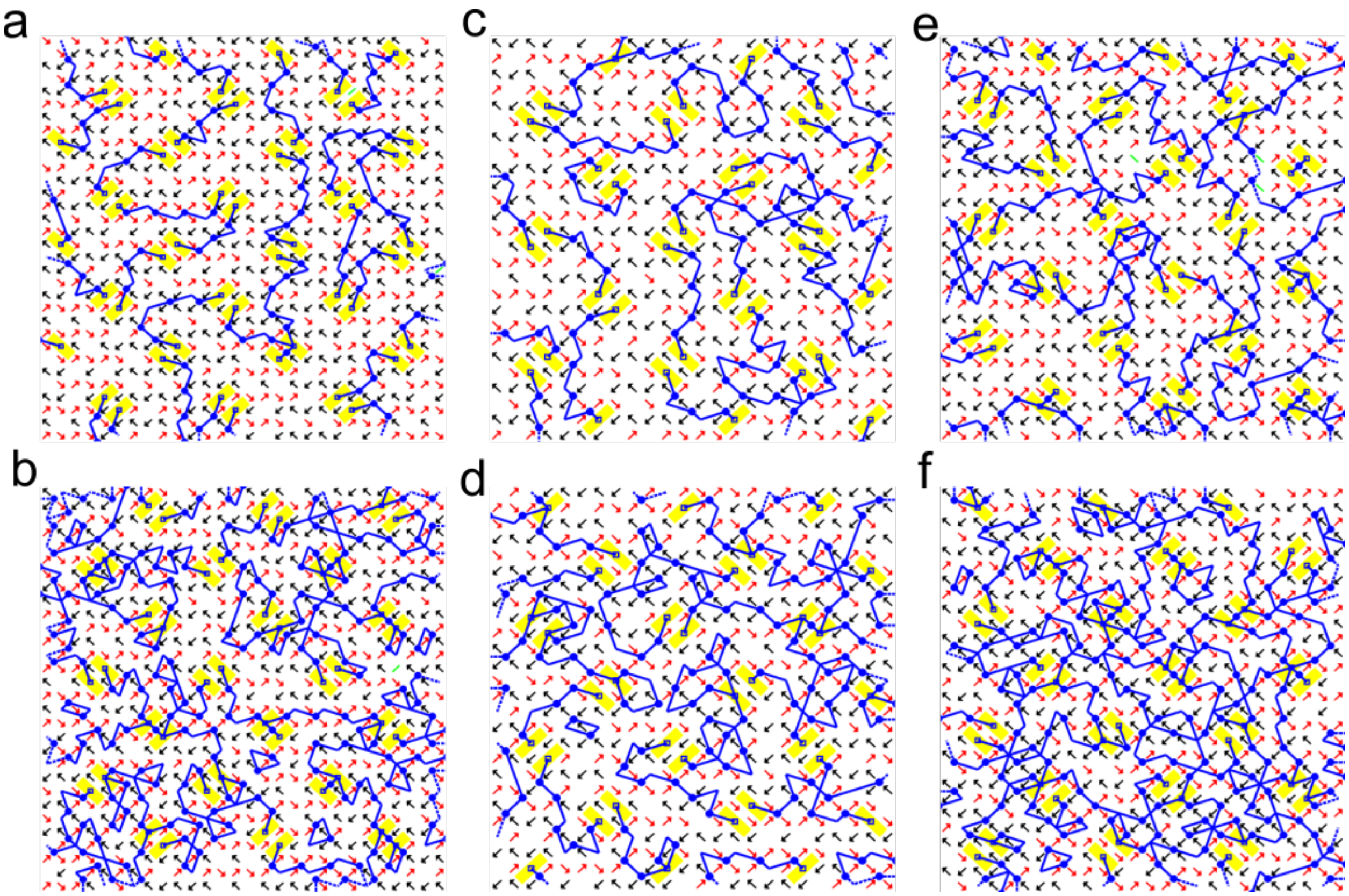

*Supplementary Fig. 11. Examples of digitized PEEM images with both the moments and strings represented for SFI of a = 600 nm at (a) 250 K and (b) 350 K, 700 nm at (c) 250 K and (d) 350 K, and 800 nm at (e) 250 K and (f) 350 K. Broken lattice lines indicate that the islands were outside of the PEEM image boundary. Green lines indicate that the moment directions were not resolvable from PEEM images. Dashed blue lines give the possible string configurations around missing moments.*

**Supplementary Note 5: Calculation of String Length Distribution.**

We use a simple graph theory approach to calculate the string length distribution from our experimental maps of the moments acquired from PEEM and MFM measurements. We created an adjacency matrix $\boldsymbol{A}$ for each PEEM image in which each row or column index corresponds to a vertex in the PEEM image. If the $i^{th}$ vertex and the $j^{th}$ vertex are connected by a string, the $a_{ij}$ and $a_{ji}$ entries in the matrix $\boldsymbol{A}$ are 1. Otherwise, the entries are 0. An example is shown in *Supplementary Fig. 12*. If $a_{ij}^{(L)}$ is a non-zero number in the matrix $\boldsymbol{A}^{L}$, the $i^{th}$ vertex and the $j^{th}$ vertex are connected by a string of length $L$. We calculate the matrix $\boldsymbol{A}^{L}$ for integral $L$ = 1 to $N$ where $N$ is the maximum string length. The number of strings of length $L$ comes from the number of entries $a_{ij}^{(L)} > 0$ ($i<j$). Note that $a_{ij}^{(L)}$ must satisfy three conditions: (1) both the $i^{th}$ vertex and the $j^{th}$ vertex have a string that directly go into an interior plaquette, (2) $a_{ij}^{(m)} = 0$ for all $m<L$, and (3) there is no such index $t$ (the $t^{th}$ vertex is also directly connect to an interior plaquette), where $a_{it}^{(p)} > 0$, $a_{tj}^{(q)} > 0$, and $p+q=L$. The first condition guarantees that we only look at the strings connecting interior plaquettes. The second condition makes sure every segment on a string is counted only once. And the third condition excludes the strings that go from interior plaquettes A to B through C.

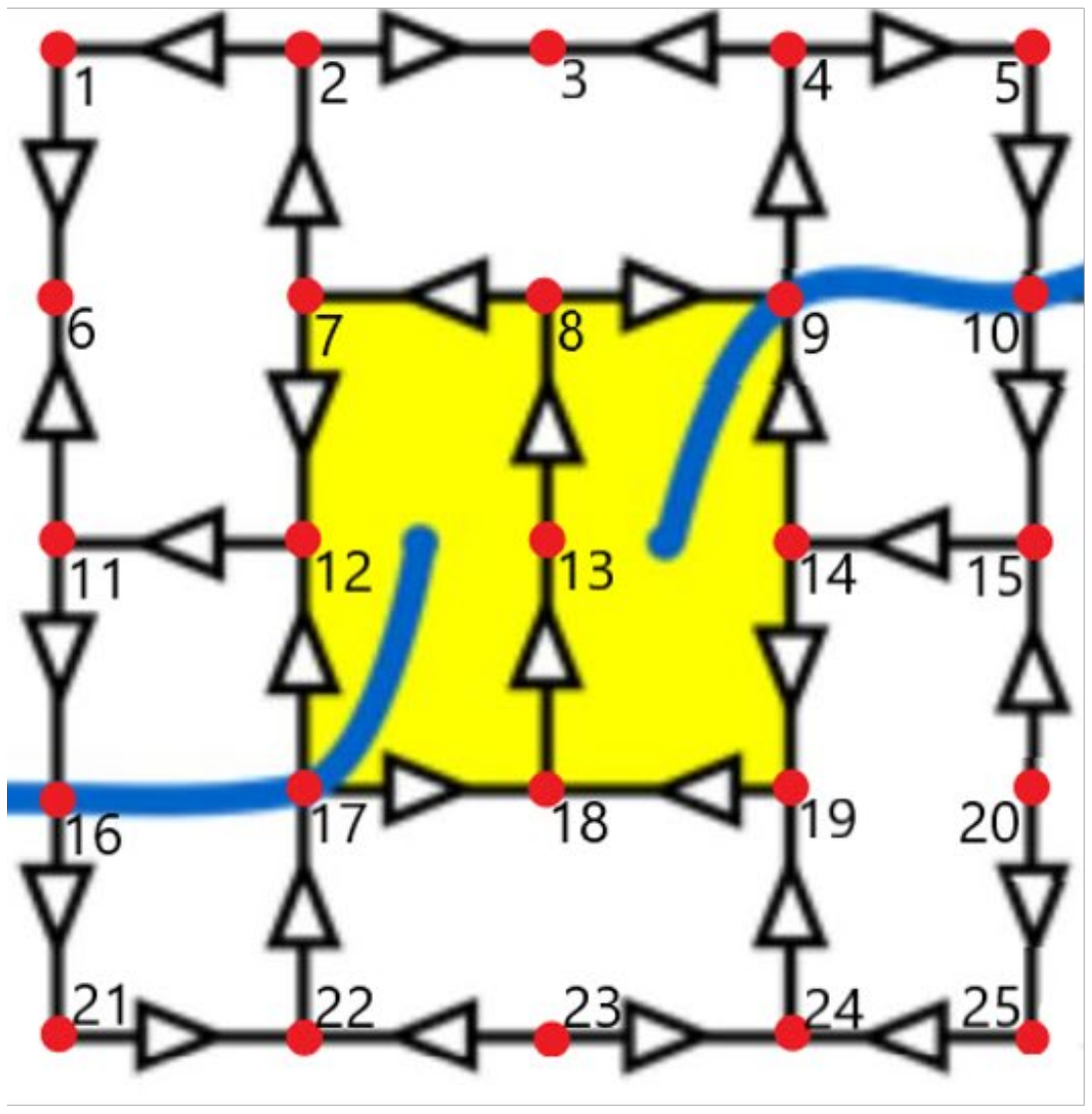

| | 1 | 2 | ... | 9 | 10 | ... | 16 | 17 | ... | 25 |
|---|---|---|---|---|---|---|---|---|---|---|
| 1 | 0 | 0 | ... | 0 | 0 | ... | 0 | 0 | ... | 0 |
| 2 | 0 | 0 | ... | 0 | 0 | ... | 0 | 0 | ... | 0 |
| ... | 0 | 0 | ... | 0 | 0 | ... | 0 | 0 | ... | 0 |
| 9 | 0 | 0 | ... | 0 | 1 | ... | 0 | 0 | ... | 0 |
| 10 | 0 | 0 | ... | 1 | 0 | ... | 0 | 0 | ... | 0 |
| ... | 0 | 0 | ... | 0 | 0 | ... | 0 | 0 | ... | 0 |
| 16 | 0 | 0 | ... | 0 | 0 | ... | 0 | 1 | ... | 0 |
| 17 | 0 | 0 | ... | 0 | 0 | ... | 1 | 0 | ... | 0 |
| ... | 0 | 0 | ... | 0 | 0 | ... | 0 | 0 | ... | 0 |
| 25 | 0 | 0 | ... | 0 | 0 | ... | 0 | 0 | ... | 0 |

Adjacency matrix A

*Supplementary Fig. 12. An example of a spin map and its adjacency matrix* ***A****. The 9$^{th}$ and 10$^{th}$ vertices and 16$^{th}$ and 17$^{th}$ vertices are connected by strings. Therefore, the* $a_{9,10}$, $a_{10,9}$, $a_{16,17}$, *and* $a_{17,16}$ *are 1, marked in red. Other entries are 0.*

## SUPPLEMENTARY REFERENCES